\documentclass[twocolumn,samsmath,amssymb,floatfix]{revtex4}
\usepackage{graphicx}
\usepackage{dcolumn}
\begin{document}

\title{Density fluctuations in the intermediate glass-former glycerol:\\
a Brillouin light scattering study}

\author{
    Lucia Comez, Daniele Fioretto, Filippo Scarponi}
\address{
     Dipartimento di Fisica and INFM, Universit\`a di Perugia,
     I-06123, Perugia, Italy.}

\author{Giulio Monaco}

\address{European Syncrotron Radiation Facility, B.P. 220 F-38043,
Grenoble, France.}

\begin{abstract}
Brillouin scattering has been used to measure the dynamic structure
factor of glycerol as a function of temperature from the high
temperature liquid to the glassy state. Our investigation aims
at understanding the number and the nature of the relaxation
processes active in this prototype glass forming system in the high
frequency region. The associated character of glycerol is reflected
by a rather simple relaxations pattern, while the
contributions coming from intra-molecular channels are
negligible in the GHz frequency region. The temperature behavior
of the characteristic frequency and lifetime of the longitudinal acoustic
modes is analyzed, suggesting that a phenomenological model
which only includes the structural ($\alpha$) process and the
unrelaxed viscosity is able to catch the leading contributions to
the dynamics of the density fluctuations. This ansatz is also
supported by a combined analysis of light and inelastic x-ray
scattering spectra. The temperature dependence of the
characteristic time of the $\alpha$-process, $\tau_\alpha$, obtained
by a full-spectrum analysis conforms to the $\alpha$-scale universality,
i.e. the values $\tau_\alpha$ revealed by different experimental
techniques are proportional the ones to the others.
The non-erogodicity parameter smoothly decreases on increasing the
temperature, and no signature of the cusp-like behavior predicted by
the idealized mode coupling theory and observed in other glass-formers
is found in glycerol.
\end{abstract}


\maketitle

\section{Introduction}
\label{intro}

Density fluctuations hold a key role in the dynamics of
glass-forming systems: the glass formation itself is related to
the progressive arrest of long time density fluctuations. The
characteristic time $\tau_{\alpha}$ of these fluctuations
exhibits a steep increase for temperatures approaching the glass
transition temperature, $T_g$, ranging from $\sim$10$^{-12}$ s
well above the melting point to $\sim$10$^{2}$ s at $T_g$. For
this reason more than one technique is usually required to cover
the whole time or frequency range where this dynamics takes place.
Inelastic neutron, x-ray and light scattering are the experimental
techniques most widely used  to study the density fluctuations in
supercooled systems. Among these, Brillouin light scattering
(BLS) is a powerful tool to gain access to the spectrum of density
fluctuations in the GHz region.

In the earliest BLS studies of glass forming systems, the data
analysis was focused on measuring the frequency shift and
linewidth of the Brillouin peaks, i.e. the phase velocity and
absorption of the longitudinal acoustic modes
\cite{Wang,Torell,Patterson84}. The analysis of the temperature
dependence of these parameters was used to determine either the
temperature dependence of the relaxation time (activation plot)
once the shape of the relaxation function was fixed, or,
conversely, the shape of the structural relaxation (master plot)
once the $\tau$(T) behavior was given \cite{Floudas1991}.

More recently, the improvements of interferometric setups allowed
the experimentalists to obtain high-contrast and high-resolution
BLS spectra in a wider frequency window ranging from few hundreds
of MHz to few tens of GHz, thus gaining information on the shape
of both the Brillouin peaks and the quasi-elastic central region
known as Mountain peak \cite{mountain}. Thanks to these advances,
it has become possible to carry out full spectrum analyses in a
window of 1$\div$2 decades in frequency, using expressions for
the spectrum of density fluctuations derived within the
Generalized Hydrodynamics formalism \cite{Montrose} or,
equivalently, from a generalized Langevin equation \cite{boon}.
The spectrum is thus written in terms of static correlators and
of a frequency-dependent longitudinal modulus or, equivalently,
longitudinal viscosity. In the earliest investigations of
supercooled systems, reasonable fits of the measured spectra were
obtained considering the $\alpha$-relaxation as the only relevant
relaxational process. In those investigations, the
$\alpha$-relaxation was typically described by the Cole-Davidson
(CD) expression, like in the case of salol \cite{Dreyfus1992}, PC
\cite{Torell1990}, ZnCl$_2$ \cite{Dreyfus2001,Soltwisch1987}. The
main limit of this approach is to ignore the possible presence of
additional processes in the BLS frequency range, faster than the
CD relaxation.

A contribution to the fast dynamics of density fluctuations in
supercooled systems is predicted by the mode coupling theory (MCT)
\cite{mct} as the fast component of the structural relaxation,
also known as the mode coupling $\beta$ region. During the last
decade an important contribution to the comparison of the MCT
predictions to Brillouin spectra has been given by Cummins and
coworkers \cite{Cumminstot}. Another important contribution to
the fast dynamics of density fluctuations in molecular liquids
comes from internal thermal relaxations \cite{herzfeld}. These
are connected to the exchange of energy between acoustic waves
and intramolecular degrees of freedom, like vibrations and
rotations of groups of atoms. It is interesting to notice that
the Generalized Hydrodynamic approach to Brillouin scattering was
originally introduced in the sixties by Mountain and Zwanzig
\cite{mountain,Zwanzig1965} to explain exactly the thermal
relaxation in Kneser-type liquids \cite{herzfeld}, although its
importance has been largely underestimated in more recent studies
of supercooled systems. A clear signature of the existence of
this relaxation channel in BLS spectra can be found in the almost
Arrhenius temperature dependence of the relaxation time obtained
by the single-relaxation analysis of different glass forming
systems, both simple \cite{Torell1990} and polymeric (see for
instance, \cite{Firettoold,Levelut}) in nature.

The main reason why a two-relaxation
($\alpha$+thermal) model was not employed for long time in the
analysis of BLS spectra can be found in the narrowness of the
frequency window available to this technique. A step forward in the
solution of this problem has been recently proposed, based on the
combined use of ultrasonic, BLS and inelastic x-ray scattering (IXS)
spectra. IXS is a novel technique which directly probes
density fluctuations in the high q limit of some
nm$^{-1}$, extending the frequency range accessible to Brillouin
scattering investigations to the THz region. It is then now possible to
carry out a full spectrum analysis in a reasonably wide spectral
region, including both structural, thermal and unrelaxed viscosity
("instantaneous") processes. Using this method in polybutadiene
\cite{fiorettoPRB02} it was possible to separate the contribution
of the $\alpha$ relaxation from that of the fast intramolecular
process, and to reconcile the positive tests of MCT previously
obtained by neutron scattering with BLS results. In
ortho-terphenyl, the presence of a vibrational relaxation
\cite{otpfast1,otpfast2} has been confirmed by BLS measurements
on the single crystal \cite{otpcri}, and by molecular dynamics
simulations \cite{mossa}.

In this work, we study the high frequency dynamics of the glass
former glycerol, an associated liquid which forms a H-bond
network, making it intermediate between fragile and strong
systems. Different from Kneser liquids, in associated liquids the
ratio between the acoustic absorption $\alpha$ and the classical
absorption $\alpha_{cl}$ (only due to shear viscosity and thermal
conductivity) is very small (typically less than 3) and
temperature independent. This feature is traditionally attributed
to a negligible coupling of the acoustic waves with internal
degrees of freedom. This simple relaxation pattern suggests
glycerol as a natural candidate for testing glass formation
theories.

Glycerol has already been the object of several studies concerning
its acoustic properties and dynamical
processes. Ultrasonic (US) measurements in the MHz frequency
region have been reported of the adiabatic speed and of the absorption
of sound \cite{Picci57,Meis60,herzfeld,NagelPRA86}. Photon
correlation spectroscopy (PCS) has been used to characterize the
structural relaxation close to $T_g$ \cite{Demoulin74,Dux79}.
Impulsive stimulated Brillouin and thermal light scattering (ISBS
and ISTS) experiments \cite{YanChen88,Nelson00} have provided
information on the sound dispersion and absorption and on the
structural relaxation at low temperatures and for
different $q$ values. Stimulated Brillouin gain (SBG)
spectroscopy has been also used to probe the dynamics of
supercooled and glassy glycerol \cite{TandyGrubbs94}. Very
recently, the density response of supercooled glycerol to an
impulsive stimulated thermal grating has been studied in the
200-340 K temperature range in order to separate the structural
relaxation from thermal diffusion processes \cite{DiLeo02}.
Rayleigh-Brillouin spectra at very high resolution have been
presented \cite{vacher} to reveal acoustic anomalies in glassy glycerol
at low temperatures. The temperature
behavior of the structural relaxation process has also been studied by
depolarized light and neutron scattering
\cite{roesslerPRL94,wuttkePRL94,sokolov95}, and dielectric
spectroscopy \cite{schonhalsPRL93,LunkPRL96,Schneider98}.
Moreover, inelastic x-ray scattering
\cite{Sette98,Masciove98,Ruocco99} has been used to investigate
the high-frequency dynamics of supercooled glycerol and
the nature of the acoustic attenuation at low temperatures.
Despite of this extensive literature, there is no light scattering
investigation which covers the entire range comprised between the
high temperature liquid and the glassy state.

Here we report high resolution Brillouin spectra covering nearly
two decades in frequency in a wide temperature range from 47 to
441 K. The main purpose of this paper is to contribute to the
identification of the relaxation processes active in the GHz
frequency region of glycerol, and to the characterization of their
temperature evolution. In order to constrain the hydrodynamic model
proposed for the density fluctuations, we perform a combined light
and inelastic x-ray analysis. Moreover, a comparison is also reported
with the results obtained by different techniques, including ISTS,
ISBS, SBG, dielectric spectroscopy, photon correlation spectroscopy,
and depolarized light scattering, in order to test the degree of
universality of the temperature dependence of the characteristic time
and of the shape parameters of the $\alpha$ relaxation.

Finally, a preliminary comparison with the predictions of the
mode coupling theory of supercooled liquids is presented. MCT
\cite{mct}, considered the first microscopic model of the glass
transition, describes the slowing down of the structural dynamics
in terms of a non-linear coupling between density fluctuations
which, if we neglect hopping processes, causes the structural
arrest of the system at the critical temperature, $T_c$. Several
studies have been previously performed to test the MCT
predictions on glycerol. Depolarized light scattering
\cite{roesslerPRL94,wuttkePRL94,sokolov95} and neutron
\cite{wuttkePRL94} and dielectric
\cite{schonhalsPRL93,LunkPRL96,Schneider98,roesslerJPCM03} data
have been subjected to MCT checks. All together these studies
provide a variegated picture of the existence and the eventual
location of the critical temperature, $T_c$: the different $T_c$
evaluations span a range comprised between 250 and 310 K
\cite{roesslerPRL94,wuttkePRL94,sokolov95,LunkPRL96,Schneider98,schonhalsPRL93,roesslerJPCM03}.

One of the main predictions of the idealized mode coupling theory
concerns the
temperature behavior of the non-ergodicity factor $f_q$ --- the
normalized amplitude of the structural relaxation --- which
should exhibit a cusp at $T_c$. In this paper, by measuring the
limiting low and high frequency adiabatic sound velocity in a wide
temperature range, we derive the non-ergodicity factor
\cite{goetzeLONG} for the density fluctuations, and look for the
possible existence of the crossover temperature predicted by MCT.

\section{Theoretical background}
\label{theory}

\subsection{Dynamic structure factor}
\label{model}

An expression for the dynamic structure factor, and thus for
isotropic Brillouin spectrum which is proportional to it, can
be derived from the equation of motion of the microscopic density
$\rho(r,t)$. In the limit of low thermal conductivity, density
fluctuations are decoupled from temperature fluctuations, and
the equation of motion of the $q$-th component of the microscopic
density, $\rho(q,t)$, is derived within simple hydrodynamics and
comes out to be that of a damped harmonic oscillator (DHO):

\begin{equation}\label{eqmoto}
\left[ \left( \frac{\rho_M}{q^2}\right) \frac{\partial^2}{\partial
t^2}+\eta_L\frac{\partial}{\partial t}+ M \right] \rho (q,t)=0
\end{equation}

where $\rho_M$ is the average mass density, $\eta_L$ the
longitudinal viscosity and $M$ the longitudinal elastic modulus.
The dynamic structure factor, $S(q,\omega)$, obtained from the
Laplace-Fourier transform of Eq.~(\ref{eqmoto}), is:

\begin{equation}\label{Sqeomega}
S(q,\omega)=\frac{S(q) M}{\pi \omega}\frac{\omega \eta_L
}{[\omega^2\rho_M/q^2-M]^2+ [\omega \eta_L]^2},
\end{equation}
where $S(q)$ is the static structure factor. A Brillouin peak in
the spectrum is thus expected close to

\begin{equation}
\omega_{LA}=q(M/\rho_M)^{1/2},
\end{equation}
proportional to the longitudinal acoustic (LA) velocity
$c_{LA}=\omega_{LA}/q$. The full width of the peak

\begin{equation}
\Gamma_{LA}=q^2 \eta_L /\rho_M
\end{equation}
is proportional to the attenuation of the LA mode. In fact, the
absorption of longitudinal acoustic modes is given by
${\alpha}=\Gamma_{LA}/2 c_{LA}$.
The integrated area of the spectrum is proportional to the
adiabatic compressibility $\chi_S$. In the limit of low thermal
conductivity here considered, thermal fluctuations contribute only
to the low frequency region of the spectrum, giving rise to the
Rayleigh peak centered at frequency zero and linewidth of about
0.1 GHz. The total intensity, taking into account also this last
contribution, is proportional to the isothermal compressibility
$\chi_T$. The ratio of the intensity of the Rayleigh line over
the Brillouin lines $(\chi_T - \chi_S)/ \chi_S = \gamma-1 =
R_{LP}$ is known as Landau-Placzek ratio.

In molecular liquids the acoustic modes may couple both with the molecular
internal degrees of freedom and with the $\alpha$ process, giving rise to
additional mechanisms of acoustic damping. These relaxation
effects are also responsible for a peak at zero-frequency (Mountain
peak) and can be taken into account by a generalization of the
hydrodynamic equations, introducing an $\omega$-dependent modulus
(or viscosity). In particular, the complex frequency dependent modulus
$M(\omega)=M'(\omega)-iM''(\omega)$ can be written as:

\begin{equation}\label{modulus}
M(\omega)=M_\infty+\Delta M(\omega)+i\omega \eta_\infty,
\end{equation}
where $M_\infty$ is the high-frequency (unrelaxed) longitudinal
modulus and $\eta_\infty$ is the high-frequency longitudinal
viscosity which accounts for relaxation processes characterized
by a time so short to be out of the experimental frequency
window. In the low frequency (relaxed) limit, $\Delta
M(\omega\rightarrow0) = (M_0 - M_\infty) + i\omega (\eta_0 -
\eta_\infty)$, where $M_0$ ($\eta_0$) is the zero-frequency
longitudinal modulus (viscosity). In this limit, the complex
longitudinal modulus becomes $M(\omega \rightarrow 0)= M_0 + i
\omega \eta_0$. This is the previously described simple
hydrodynamics regime, where $\eta_0\equiv\eta_L$. In the opposite
limit, the unrelaxed one, $\Delta M(\omega \rightarrow \infty)
=$0 and the longitudinal modulus takes the simple form:
$M(\omega) = M_\infty + i\omega \eta_\infty$. This coincides with
the Voigt model of viscoelasticity which is used since long time
to describe the acoustic damping in solids (see, for instance,
\cite{philippoff}). In this limit the spectrum takes the same
form of Eq.~(\ref{Sqeomega}), but with $M$ replaced by $M_\infty$
and $\eta_L$ by $\eta_\infty$. The intermediate condition between
the relaxed and unrelaxed limits is the one where the relaxation
processes strongly couple with the acoustic waves, and the
detailed description of the shape of the Brillouin spectrum
requires the knowledge of the frequency dependent part of the
modulus, $\Delta M(\omega)$. In this condition
Eq.~(\ref{Sqeomega}) can still be used to fit the spectrum in a
narrow frequency region around the Brillouin peaks to obtain
values for the "apparent" sound velocity and attenuation (see for
instance, \cite{otpfast2,fiorettoepon}), i.e. their values at the
frequency of the peak. This is the "traditional" way of analyzing
Brillouin spectra, safe enough also for measurements obtained with
low contrast and non-tandem interferometers. The analysis of the
temperature behavior of these parameters gives useful information
on the number and typology of relaxation processes active in the
Brillouin frequency window, and will be adopted in the first part
of the data analysis, paragraph \ref{short}.

In the most general case of coupling of relaxation processes with
density fluctuations, the dynamic structure factor measured with both
the BLS and the IXS techniques can be expressed in terms of
the generalized modulus of Eq.~(\ref{modulus}):

\begin{equation}\label{Iroro}
S(q,\omega)=\frac{S(q) M}{\pi \omega}\frac{M''(\omega)
}{[\omega^2\rho_M/q^2-M'(\omega)]^2+ [M''(\omega)]^2}.
\end{equation}

In this case, the simplest possible situation is that of an exponential
relaxation described by the Debye formula
$\Delta M(\omega)=(M_0 - M_\infty)/(1+i\omega
\tau)$, where $\tau$ is the relaxation time. Notice that $M_0 <
M_\infty$, i.e. the modulus increases with increasing frequency.
The imaginary part of the modulus shows a peak at
$\omega=\tau^{-1}$ corresponding to a maximum of acoustic energy
absorption. It occurs, for instance,  when the frequency
of the acoustic mode matches the rate of energy
exchange with some internal degree of freedom.  The spectral
width of the peak of $M''(\omega)$ for a Debye relaxation is of
1.14 decades, in the same frequency region where the corresponding
real part changes from $M_0$ to $M_\infty$ (dispersion). Viscoelastic
systems approaching the glass transition are characterized by a
dispersion which extends in a much wider frequency range (2 or 3
decades is not a rare case). This stretching phenomenon is frequently
accompanied by a power law dependence of $M''(\omega)$ which can
be described by the Cole-Davidson relaxation function

\begin{equation}\label{CD}
\Delta M(\omega)=(M_0 - M_\infty)/(1+i\omega \tau)^\beta,
\end{equation}
where $\beta$ is the stretching parameter, being
$M''(\omega)\propto \omega^{-\beta}$ on the high frequency side
of the peak in $M''(\omega)$.
Fitting Brillouin spectra with this phenomenological ansatz for
the longitudinal modulus gives the values of the amplitude $(M_0
- M_\infty)$, relaxation time and stretching parameter at each
measured temperature. The study of the temperature behavior of
the $\alpha$-relaxation parameters is the main topic of the
physics of the glass transition. This is the content of
Sec.~\ref{disc}.

No mention has been given up to now of the microscopic origin of
the $\alpha$-relaxation. MCT suggests that in the liquid above
$T_g$, the non-linear coupling of density fluctuations
characterized by different values of $q$ is the mechanism
responsible for it. This theory gives a closed set of equations
relating the dynamics of density fluctuations to the static
structure factor of the same liquid. A full check of the
predictions of MCT is beyond the aims of the present work. We
limit ourselves to note that the Cole-Davidson relaxation
function is usually adequate to represent the slow part of the
relaxation function as obtained by the MCT \cite{mct} and its
amplitude will be used in Section~\ref{DW} to check the existence
of the square root singularity in the non-ergodicity factor, as
predicted by the theory \cite{mct,goetzeLONG}.

\bigskip

\section{Experimental details}
\label{expt}

\subsection{Brillouin light scattering measurements}
\label{exBLS}

\begin{figure}
\hspace{-10.0cm}{\scalebox{1.2} {\includegraphics{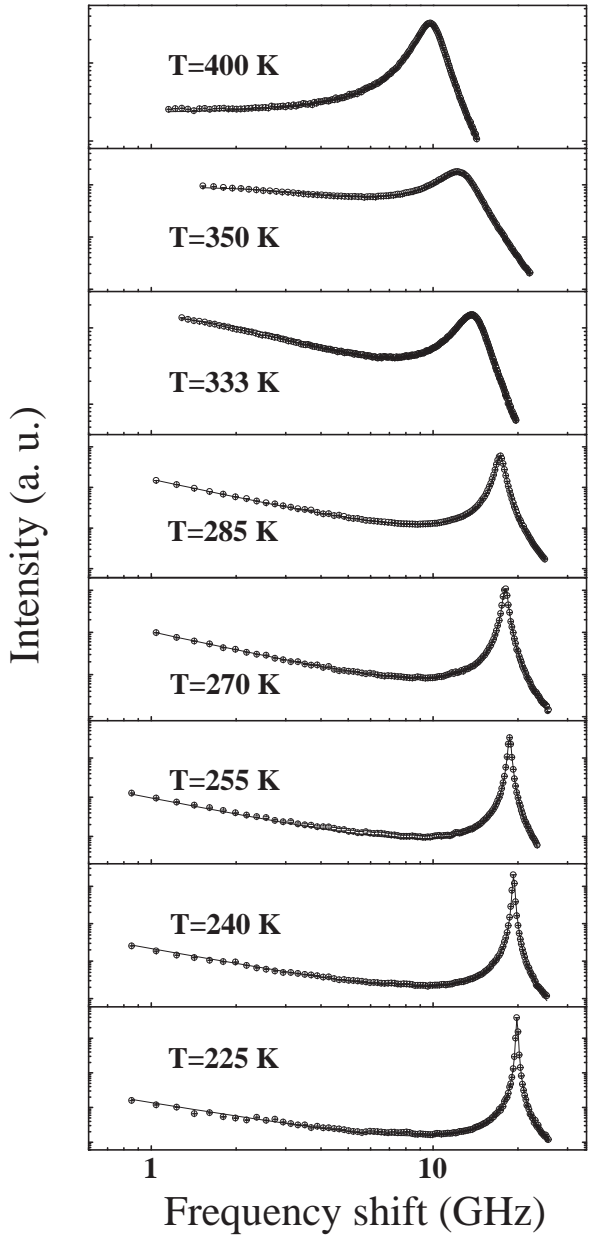}}}
\caption{\label{spettri} Isotropic Brillouin light scattering
spectra as a function of temperature for normal and supercooled
liquid glycerol. The lines through the experimental points are
the best fits obtained using Eqs.~6 and 14.}
\end{figure}

Glycerol of 99.5\% purity purchased from Aldrich Chemicals has
been employed.  The sample was filtered through a 0.22 $\mu m$
teflon filter directly into the cylindrical pyrex cell (inner
diameter $\approx $ 1 cm) used for the measurements and slowly
degassed under vacuum. The cell was sealed without breaking the
vacuum. Brillouin spectra were collected in a wide temperature
range comprising the whole liquid - normal and supercooled -
range (melting point, $T_m=$ 291 K) down to the glassy state
($T_g=$ 187 K). The sample was placed in a copper holder which
was used to regulate the temperature. For the measurements above
room temperature, the sample holder was heated with a NiCr wire
using a conventional dc power supply. The temperature was
controlled by a Thor Cryogenics 3010 II model temperature
controller and measured by a Pt100 calibrated resistance placed
close to the sample. The measurements were started after the
sample temperature was thoroughly equilibrated and the
temperature fluctuations were kept within $\pm 0.1$ K during the
spectra recording. For the low-temperature measurements a
Cryomech ST405 cryostat was used. The sample temperature, varied
with a typical 0.5 degree/min rate, was measured with a 410-NN-1
Silicon Diode Temperature Sensor. A Coherent-Innova 300 model
Ar$^{+}$ laser was operated with a typical power of $\approx$300
mW on a single mode of the $\lambda _o$=514.5 nm line. The light
scattered by the sample was analyzed using a Sandercock-type
(3+3)-pass tandem Fabry-Perot interferometer characterized by a
finesse of about 100 and a contrast $>$ 5$\cdot 10^{10}$
\cite{sand,ghost}. The elastic scattering from a dilute water
suspension of Latex particles (120 nm diameter) was used to
determine the instrumental resolution function $R(\omega )$. In
the first part of our work, we aimed at the determination of the
sound speed and attenuation of longitudinal acoustic modes. For
this purpose the peak position and the linewidth (FWHM) of the
Brillouin components are the only relevant parameters, making it
possible to focus on the frequency region of the Brillouin peaks,
where the polarized ($I_{VV}$) spectra replace in good
approximation the isotropic ($I_{ISO}$) ones.  In this spirit,
polarized Brillouin spectra were collected in the back-scattering
$(\theta \approx 180^o)$ geometry in a wide temperature range,
namely 47-441 K, using a narrow $T$-grid (about 5 K above room
temperature, 10 K below room temperature). Different Free
Spectral Range (FSR) values were used for the measurements above
and below room temperature (20 GHz and 10$\div $8.3 GHz,
respectively). The integration time was of $\approx$10 s /
channel and the effect of the dark counts ($<$ 1 count / s) was
checked to be negligible. Typical polarized spectra covered the
0.8$\div $25 GHz frequency range. In a successive step we
selected some temperatures above $T_g$ for a detailed analysis of
the shape of the relaxation function. We collected in the 210-400
K temperature range (namely, 400, 350, 333, 326, 295, 285, 270,
255, 240, 225, and 210 K) polarized ($I_{VV}$) and depolarized
($I_{VH}$) spectra using two different FSRs of 13.6 GHz and 50.0
GHz in order to cover the 0.8-100 GHz frequency range. Each
spectrum (polarized and depolarized) results from the overlap of
the two spectra measured with different FSRs. The isotropic
spectra $I_{ISO}$ have been then obtained by subtraction of the
$I_{VV}$ and $I_{VH}$ spectra following the procedure described in
Ref.~\cite{fiorettoepon}. Typical $I_{ISO}$ spectra are presented
in Fig.~\ref{spettri} at some selected temperatures.

\subsection{Brillouin inelastic X-ray scattering measurements}
\label{exIXS}

\begin{figure}
\scalebox{1.0} {\includegraphics{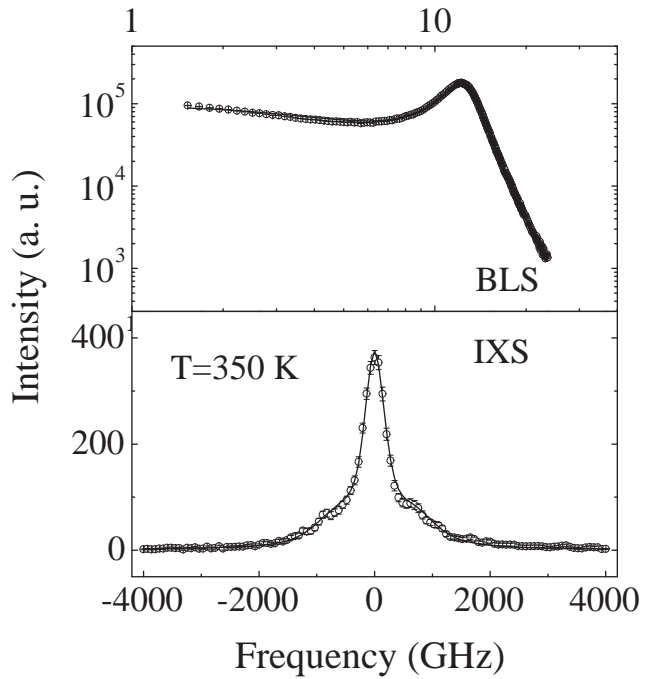}}
\caption{\label{spetIXS} A typical IXS spectrum (bottom panel)
acquired at T=350 K  and $q$=2 nm$^{-1}$ shown together with the
BLS spectrum at the same temperature (top panel). The lines
through the experimental points are the best fits obtained from
the combined BLS-IXS analysis using Eqs.~6 and 14.}
\end{figure}

Inelastic X-ray measurements were carried out at the very high
energy resolution IXS beamline (ID16) of the European Syncrotron
Radiation Facility \cite{esrfID16} using the Si(11 11 11)
reflection for both the monocromator and the analyzers. The total
energy resolution is of $\approx$ 1.5 meV. The IXS spectra are
directly proportional to $S(q,\omega)$. IXS measurements have
been performed in the 295-400 K temperature range, for $q$ values
in 2-11 nm$^{-1}$ interval, with a $q$-resolution of about 0.4
nm$^{-1}$. Experimental data were normalized to the intensity of
the incident beam and the contribution of the empty cell was
subtracted from each spectrum. A typical IXS spectrum acquired at
T=350 K  and $q$=2 nm$^{-1}$ is shown in Fig.~\ref{spetIXS}
together with the BLS spectrum at the same temperature.

\subsection{Ultrasonic measurements}
\label{us}

The temperature behavior of the relaxed sound velocity $c_0$ was
measured by an ultrasonic technique at 5 MHz \cite{fiorettoIEEE},
i.e. at a frequency four orders of magnitude lower than that
corresponding to the LA modes revealed by Brillouin light
scattering. A PZT transducer was placed deep into the sample, at
a distance $L=5$mm from the bottom of a brass cell parallel to
the surface of the transducer. The system as a whole works as an
acoustic Fabry-Perot, so that the reflection coefficient, $R$,
shows a series of minima corresponding to the condition $L=m
\Lambda/2$, where $\Lambda$ is the acoustic wavelength and $m$ is
an integer number. The reflection coefficient is measured as a
function of frequency by means of an HP8753A Network Analyzer.
The distance $\Delta \nu$ between two neighboring minima, the
free spectral range, is thus given by the relation $\Delta
\nu=c/2L$, and the Fourier transform of $R$ shows a well defined
maximum at $t=2L/c$, where $c$ is the velocity of the acoustic
wave in the sample. From the measurements of $t$, the velocity
$c$ of longitudinal acoustic waves was obtained at several
temperatures in the range 306.5-389.5 K.

\section{Data analysis}
\label{analysis}

\subsection{Acoustic properties of glycerol}
\label{short}

The first step of our investigation is a traditional acoustic
analysis of the Brillouin spectra consisting in the measurement
of the frequency position and linewidth of the Brillouin peaks as a
function of temperature. It has been shown in Sec.~\ref{model}
that the isotropic Brillouin spectrum, for frequencies close to
those corresponding to the longitudinal acoustic (LA) peaks, can
be represented by a damped harmonic oscillator (DHO) model:

\begin{equation}\label{dhopicco}
I_{LA}(\omega)={I_{LA}^0}\frac{\Gamma_{LA}\omega^2_{LA}}{[\omega^2_{LA}-\omega^2]^2+[\omega\Gamma_{LA}]^2},
\end{equation}
where $\omega_{LA}/2\pi$ and $\Gamma_{LA}/2\pi$ correspond to the
frequency position and to the full width at half-maximum (FWHM)
of the LA peaks.

\begin{figure}
\hspace{2.9cm} \vspace{0.0cm}{ \scalebox{0.8}
{\includegraphics{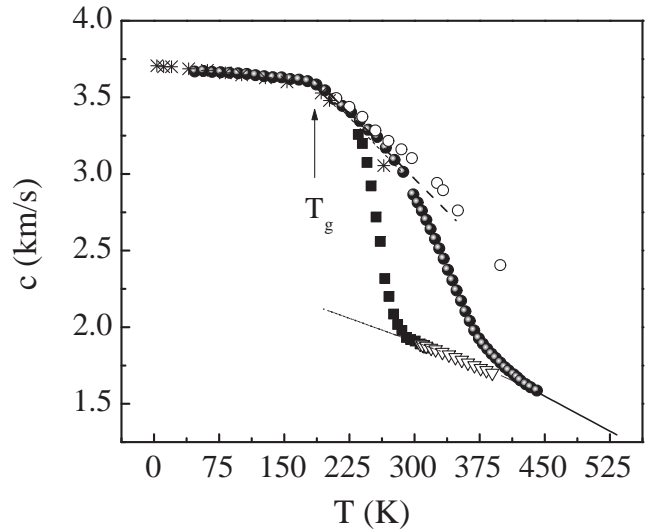}}} \caption{\label{velbril} Sound
velocity data from the present Brillouin light scattering
experiment (full circles) compared with literature data: stars
represent previous BLS measurements from \cite{vacher}, full
squares represent ultrasonic measurements at 15 MHz from
\cite{Picci57}. Limiting high frequency values obtained from our
BLS full spectrum analysis are reported as open circles, together
with the expressions given in
Refs.~\cite{Picci57,Meis60,YanChen88,TandyGrubbs94} (dash-dot
line). Limiting low frequency data determined from our US
measurements are reported as open triangles together with linear
expressions given in
Refs.~\cite{Picci57,Meis60,YanChen88,TandyGrubbs94} (full line).}
\end{figure}

\begin{figure}
\hspace{-0.9cm} \vspace{0.0cm}{ \scalebox{0.8}
{\includegraphics{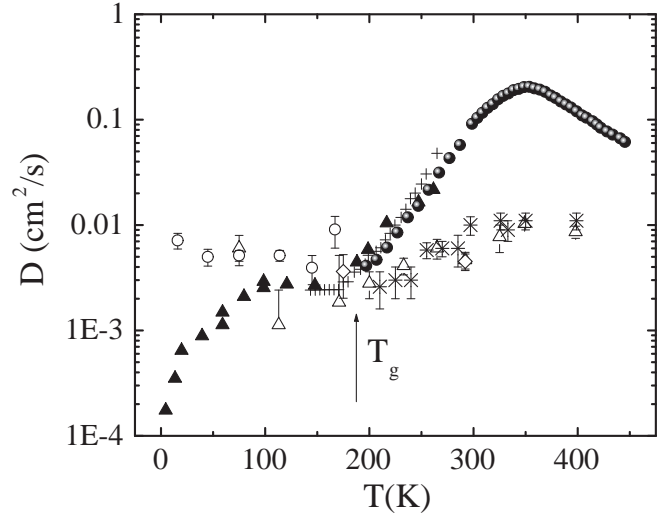}}} \caption{\label{viscocinbril}
Apparent kinematic viscosity $\Gamma_{LA}/q^2$ derived from the
analysis of the Brillouin spectrum (full circles) compared with
literature data. Crosses are from Ref.~\cite{TandyGrubbs94}, full
triangles from Ref.~\cite{vacher}. Limiting high frequency values
from Refs.~\cite{Ruocco99,Masciove98,Sette98} are reported as
diamonds, open triangles, and open circles respectively. Stars are
the unrelaxed values determined from the BLS full spectrum
analysis presented here.}
\end{figure}

The spectra collected in the 47-441 K temperature range were
fitted using Eq.~(\ref{dhopicco}) convoluted with the
experimental resolution function. The values of
$\omega_{LA}/2\pi$ and $\Gamma_{LA}/2\pi$ obtained with this
procedure are reported in Table \ref{tabni}. The values of the
(apparent) longitudinal sound velocity and of the (apparent)
longitudinal kinematic viscosity have been derived using the
relations: $c=\omega_{LA}/q$ and $D=\Gamma_{LA}/q^2$, and are
shown respectively in Fig.~\ref{velbril} and
Fig.~\ref{viscocinbril}, where they are compared with literature
data obtained from different techniques. Numerical values for $c$
and $D$ are reported as well in Table \ref{tabni}. The momentum
transfer $q$ at different temperatures has been calculated (see
Tab.~\ref{tabni}) using the refractive index $n$ and the relation
$q=2nk_i$ valid in the backscattering configuration. The
$T$-dependence of $n$ has been evaluated from that of the mass
density $\rho$ using the Clausius-Mossotti relation. Finally, the
$\rho$ values, in the temperature range here investigated, have
been obtained from the relationships:
\begin{equation}
\rho(T>T_g) = 1.272-6.55 \cdot 10^{-4} (T-273.15), \label{dens1}
\end{equation}

\begin{equation}
\rho(T<T_g) = 1.332-3.20 \cdot 10^{-4} (T-187),
\label{densbelowTg}
\end{equation}
with $T$ in K and $\rho$ in g/cm$^3$. Eq.~(\ref{dens1}) is from
Ref.~\cite{Meis60}, and Eq.~(\ref{densbelowTg}) has been obtained
using the thermal expansion coefficient of the glass at 180 K,
$\alpha_{glass}=2.4 \cdot 10^{-4}$ K $^{-1}$ \cite{herzfeld}.

\begin{table}
\caption{\label{tabni} Peak position ($\omega_{LA}/2\pi$) and FWHM
($\Gamma_{LA}/2\pi$) of the Brillouin peak reported in the whole
range of the investigated temperatures. These parameters have
been determined by fitting the spectral region around the
Brillouin peak using the DHO model, Eq.~(\ref{dhopicco}),
convoluted with the instrumental resolution. The values of FWHM
for temperatures lower than 196 K have not been reported because,
on decreasing the temperature, the linewidths become too small,
with respect to the resolution function, to be reliably
extracted. Column 2 reports the values of the exchanged momentum
$q$. Columns 4 and 6 report the longitudinal sound velocity and
the longitudinal kinematic viscosity, respectively.}
\begin{ruledtabular}
\begin{tabular}{cccccc}
$T$  &  $q$  &      $\omega_{LA}/2\pi$ &  $c$   & $\Gamma_{LA}/2\pi$  & $D$\\
$[K]$ & [nm$^{-1}$]    & [GHz]         &  [m/s] &   [GHz]        &   [cm$^2$/s]\\
\hline
& & & & &\\
47.0   &  0.0373  & 21.80  &  3668  &   ---     &  ---\\
57.0   &  0.0373  & 21.80  &  3672  &   ---     &  --- \\
67.0   &  0.0373  & 21.74  &  3666  &   ---     &  --- \\
77.0   &  0.0372  & 21.70  &  3663  &   ---     &  --- \\
87.0   &  0.0372  & 21.66  &  3659  &   ---     &  --- \\
97.0   &  0.0372  & 21.62  &  3656  &   ---     &  --- \\
107.0  &  0.0371  & 21.58  &  3652  &   ---     &  --- \\
117.0  &  0.0371  & 21.51  &  3643  &   ---     &  --- \\
127.0  &  0.0371  & 21.45  &  3637  &   ---     &  --- \\
137.0  &  0.0370  & 21.39  &  3631  &   ---     &  --- \\
147.0  &  0.0370  & 21.36  &  3630  &   ---     &  --- \\
157.0  &  0.0369  & 21.28  &  3619  &   ---     &  --- \\
166.0  &  0.0369  & 21.23  &  3614  &   ---     &  --- \\
177.0  &  0.0369  & 21.16  &  3605  &   ---     &  --- \\
187.0  &  0.0368  & 21.01  &  3583  &   ---     &  --- \\
197.0  &  0.0367  & 20.73  &  3546  &   0.09    &  0.004 \\
207.0  &  0.0366  & 20.36  &  3490  &   0.10    &  0.005 \\
217.0  &  0.0366  & 20.06  &  3444  &   0.13    &  0.006 \\
227.0  &  0.0365  & 19.77  &  3402  &   0.18    &  0.009 \\
237.0  &  0.0364  & 19.41  &  3347  &   0.25    &  0.012 \\
247.0  &  0.0364  & 19.04  &  3288  &   0.32    &  0.015 \\
257.0  &  0.0363  & 18.72  &  3241  &   0.45    &  0.022 \\
267.0  &  0.0362  & 18.28  &  3171  &   0.66    &  0.031 \\
277.0  &  0.0362  & 17.79  &  3092  &   0.90    & 0.043 \\
287.0  &  0.0361  & 17.31  &  3013  &   1.19    & 0.058 \\
298.6  &  0.0360  & 16.43  &  2866  &   1.88    & 0.091 \\
303.7  &  0.0360  & 16.11  &  2813  &   2.13    & 0.104 \\
308.6  &  0.0359  & 15.78  &  2760  &   2.38    & 0.116 \\
313.7  &  0.0359  & 15.42  &  2700  &   2.65    &  0.129 \\
318.6  &  0.0359  & 15.07  &  2640  &   2.87    &  0.140 \\
323.7  &  0.0358  & 14.68  &  2575  &   3.18    &  0.156 \\
328.6  &  0.0358  & 14.32  &  2514  &   3.43    &  0.168 \\
333.6  &  0.0358  & 13.93  &  2447  &   3.60    &  0.177 \\
338.6  &  0.0358  & 13.50  &  2374  &   3.87    &  0.190 \\
343.6  &  0.0357  & 13.10  &  2306  &   3.96    &  0.196 \\
348.6  &  0.0357  & 12.71  &  2239  &   4.12    &  0.203 \\
353.7  &  0.0357  & 12.31  &  2172  &   4.13    &  0.204 \\
358.6  &  0.0356  & 11.91  &  2103  &   3.99    &  0.198 \\
363.7  &  0.0356  & 11.54  &  2040  &   3.87    &  0.192 \\
368.8  &  0.0355  & 11.18  &  1978  &   3.68    &  0.183 \\
374.3  &  0.0355  & 10.88  &  1927  &   3.39    &  0.169 \\
378.8  &  0.0355  & 10.67  &  1892  &   3.21    &  0.148 \\
383.8  &  0.0354  & 10.48  &  1858  &   2.96    &  0.139 \\
388.8  &  0.0354  & 10.28  &  1826  &   2.77    &  0.130 \\
393.7  &  0.0353  & 10.12  &  1799  &   2.59    &  0.120 \\
399.0  &  0.0353  & 9.94   &  1769  &   2.37    &  0.110 \\
404.3  &  0.0353  & 9.75   &  1738  &   2.17    &  0.104 \\
408.5  &  0.0352  & 9.63   &  1717  &   2.06    &  0.096 \\
414.4  &  0.0352  & 9.48   &  1691  &   1.90    &  0.090 \\
418.4  &  0.0352  & 9.36   &  1672  &   1.77    &  0.084 \\
422.1  &  0.0352  & 9.24   &  1652  &   1.64    &  0.077 \\
428.2  &  0.0351  & 9.09   &  1628  &   1.51    &  0.072 \\
433.8  &  0.0351  & 8.97   &  1608  &   1.41    &  0.066 \\
441.0  &  0.0350  & 8.84   &  1586  &   1.30    &  0.061 \\
\end{tabular}
\end{ruledtabular}
\end{table}

\begin{figure}
\vspace{0.0cm} \hspace{-0.0cm}\scalebox{0.8}
{\includegraphics{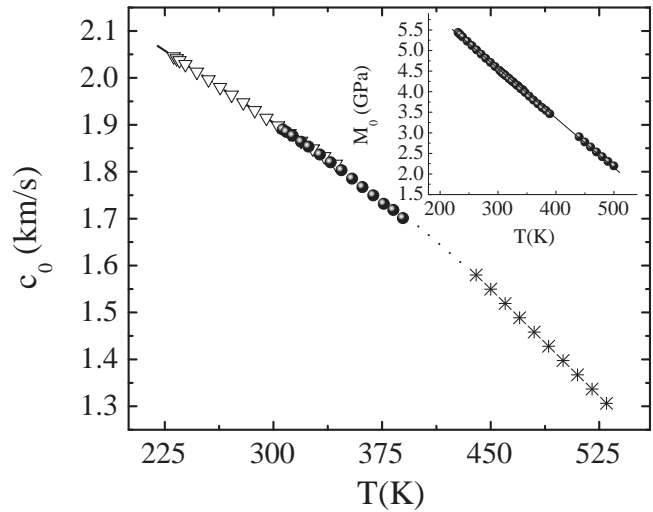}} \caption{\label{clowfreq} Relaxed
sound velocity. Our $c_0$ measurements collected at 5 MHz are
reported as full circles and are compared with previous
determinations: triangles refer to measurements reported in
Ref.~\cite{Meis60,YanChen88} and stars represent the linear
expression given in Ref.~\cite{TandyGrubbs94}. This whole set of
data does not show a linear behavior. However the same data, once
reported in terms of relaxed longitudinal modulus, $M_0$, do show
a linear behavior, as reported in the inset of the figure. The
linear representation of $M_0$ is reported in terms of $c_0$ as
dotted line.}
\end{figure}

Looking at figures \ref{velbril} and \ref{viscocinbril}, it can
be seen that our data show the well-known features expected in
presence of the structural relaxation process: a marked
dispersion of the sound velocity, accompanied by a maximum of the
absorption. In particular, the hypersonic velocity shows the
typical $S$-shape dispersion curve (full circles) that is bounded
by the limiting high (unrelaxed) and low (relaxed) frequency
values of the velocity, $c_{\infty}$ and $c_0$, respectively. The
change in the slope of $c(T)$ at about 187 K indicates the
occurrence of the liquid-glass transition. We report in the same
figure the values obtained by ultrasonic measurements of the
longitudinal modulus (squares) \cite{Picci57} showing that a
change of the probing frequency from the MHz to the GHz domain
moves the dispersion region towards higher temperatures and
broadens the temperature region where the $\alpha$ relaxation is
active. Fig.~\ref{velbril} also shows different linear
extrapolations of the unrelaxed adiabatic velocity taken from
Refs.~\cite{Meis60,YanChen88,TandyGrubbs94} (dash-dot line),
together with the values of $c_{\infty}$ obtained from the full
spectrum analysis (open circles) which will be presented in the
next section. Finally, Fig.~\ref{velbril} reports values of the
relaxed sound velocity, whose temperature behavior is essential
for the full spectrum analysis of Sec.~\ref{long}. In particular,
several sets of data for $c_0$ are available in the literature
\cite{Meis60,YanChen88,TandyGrubbs94}, obtained in different and
narrow temperature ranges below 360 K and above 440 K. They are
typically described by almost linear trends with different
slopes. In addition, we present here new ultrasonic data measured
in an intermediate temperature range between 306.5 K and 389.5 K.
These data are magnified in Fig.~\ref{clowfreq} and the
parameters for the linear approximation of $c_0$ in different
temperature regions are reported in Tab.~\ref{c0}. Once these
partial sets of data are shown all together in
Fig.~\ref{clowfreq}, a globally non-linear behavior of $c_0(T)$
becomes clear, similar to that observed in many glass-formers
\cite{harrison}. Conversely, when we plot $M_0=\rho c_0^2$, as a
function of $T$ (inset of Fig.~\ref{clowfreq}), this quantity can
be reasonably fitted by a straight line in a wide $T$-range from
245 to 505 K. The parameters of the fit are reported in
Tab.~\ref{c0}. This linear behavior of $M_0(T)$ is in agreement
with what observed by Christensen and Olsen in a restricted
temperature region \cite{Olsen94}.

\begin{table}
\caption{\label{c0} Temperature dependence of $c_0$ and $M_0$ ($T$
is in Kelvin).}
\begin{ruledtabular}
\begin{tabular}{ccccc}
$Parameter$  &  $Formula$  &  $Units$  & $T-range$ &  $Refs.$\\
\hline
& & & &\\
$c_0$  & 2519-2.051 $T$              & m/s    &  232-343 K &  \cite{Meis60}      \\
$c_0$  & 2518-2.045 $T$              & m/s    &  200-360 K &  \cite{YanChen88}    \\
$c_0$  & 2918-3.041 $T$              & m/s    &  440-534 K &  \cite{TandyGrubbs94} \\
$c_0$  & 2593-2.286 $T$              & m/s    &  307-390 K &  \footnotemark[1]      \\
$M_0$  & 8.110-1.183x10$^{-2}$ $T$   & GPa    &  245-505 K &  \footnotemark[1]       \\
\end{tabular}
\end{ruledtabular}
\footnotetext[1]{Present work.}
\end{table}

Focusing now our attention on the longitudinal kinematic
viscosity, the well defined maximum close to 350 K in
Fig.~\ref{viscocinbril} corresponds to the matching between the
relaxation rate and the characteristic frequency of the longitudinal
acoustic excitations. In the same figure we report as well the values
of $D$ determined by different techniques. Particularly relevant is the
comparison with x-ray scattering data. Generally, the IXS probing
frequency is so high that, especially for low temperatures, the
structural relaxation is definitively out of the spectral window.
Therefore, IXS typically probes the unrelaxed values of $\Gamma/q^2$,
i.e. $\eta_{\infty}/\rho$. Fig.~\ref{viscocinbril} shows that the
kinematic viscosity data obtained by BLS and IXS are consistent,
within their error-bars, in a range of about 80 K below $T_g$
beside the fact that they correspond to
$q^2$ values which differ by a factor $10^3$. Below $T_g$,
the structural relaxation is located at
frequencies lower than one Hz, and we cannot expect any
significant contribution from it to the acoustic loss in the GHz
region. The absence of an excess damping in BLS data over the IXS
unrelaxed ones at $T_g$ points in favor of an $\alpha$
relaxation-only scenario for the dynamics of glycerol. This is a
peculiar behavior of glycerol, and, possibly, of associated
liquids, different from that observed in several fragile systems
like OTP, DGEBA and PB \cite{Comez02}, where the presence of an
excess attenuation in the GHz range reflects the coupling of the LA modes
with internal, molecular degrees of freedom,
giving rise to fast relaxation processes of thermal nature. Below
100 K, the BLS values of $D$ \cite{vacher} start to decrease,
suggesting a scaling of $\Gamma$ with a power of $q$ higher than
2. A dynamic contribution to the sound attenuation having
influence on low frequency density fluctuations has been recently
proposed as the origin of this phenomenon \cite{Ruocco99,Fabian99}.

Additional information on the number and nature of the relaxation
processes active in glycerol can be obtained from the high
temperature-low frequency non dispersive region. This regime is
well described by simple hydrodynamics \cite{herzfeld}, and the
acoustic absorption, in low thermally conducting liquids, is given
by:

\begin{equation}
{\alpha}=\omega^2\frac{\eta_L}{2\rho c^3} \label{alpha}
\end{equation}
where $\eta_L=\eta_b+4/3 \eta_S$ and $\eta_L$, $\eta_b$, and
$\eta_S$ are the longitudinal, bulk and shear viscosity,
respectively. In simple liquids, where $\eta_b$ is small,
$\eta_S$ gives the most important contribution to the acoustic
attenuation, giving rise to the so-called classical absorption
\cite{herzfeld}:
\begin{equation}
{\alpha_{cl}}=\frac{2}{3} \omega^2 \frac{\eta_S}{\rho
c^3}\label{alphacl}
\end{equation}
From Eq.~(\ref{alpha}) and Eq.~(\ref{alphacl}), the relationship
between the longitudinal and the shear viscosity can be easily
written as:
\begin{equation}
\frac{\eta_L}{\eta_S}=\frac{4}{3}
\frac{\alpha}{\alpha_{cl}}\label{eta}
\end{equation}

\begin{figure}\vspace{0.0cm} \hspace{-2.0cm}{ \scalebox{0.80}
{\includegraphics{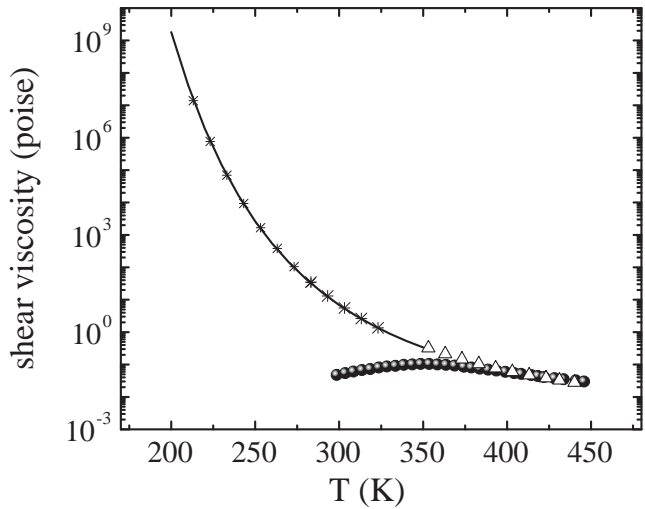}}} \caption{\label{viscobril} Shear
viscosity determined from our Brillouin data as
$\eta_S=\eta_L/2.4=\Gamma_{LA}\rho/2.4 q^2$ (full circles). These
values are compared with those obtained from direct shear
viscosity measurements: triangles refer to measurements reported
in Ref.~\cite{Vand}, stars represent measurements from
Ref.~\cite{Donth}, and the straight line indicates the expression
reported in Ref.~\cite{Meis60}.}
\end{figure}

A different temperature behavior of the $\alpha/\alpha_{cl}$ ratio
distinguishes Kneser and associated liquids \cite{herzfeld}. The
Kneser-type systems, as for example ionic liquids and the majority
of organic liquids, show a temperature dependent
$\alpha/\alpha_{cl}$ ratio ranging between 3 and 400. In these
liquids, the excess absorption over the classical value is due to
exchange of energy between internal (rotations and vibrations)
and external (acoustic modes) degrees of freedom, showing a
marked temperature dependence \cite{herzfeld}. On the contrary,
for the associated liquids, often characterized by $CH_3$, $CH_2$,
and $OH$ groups, the bulk viscosity is essentially only related
to the structural relaxation, the contributions coming from
the internal degrees of freedom being negligible. In these systems,
the acoustic measurements performed in the non-dispersive region show
a temperature independent ratio $\alpha/\alpha_{cl}$ lying between
one and three \cite{herzfeld}. In particular, for glycerol, it
has been proven that the $\alpha / \alpha_{cl}$ ratio is
approximately constant and equal to 1.78 \cite{herzfeld}.

On this basis, starting from the values of $\eta_L$ obtained from
the present BLS data, and assuming the same temperature
dependence for shear and volume viscosity, $\eta_S$ has been
obtained using the Eq.~(\ref{eta}), and $\alpha/\alpha_{cl}=1.78$
\cite{herzfeld}. In Fig.~\ref{viscobril} we report such an
evaluation for $\eta_S$ together with viscosimetric measurements
\cite{Meis60,Vand,Donth}. In the high temperature region, the two
data sets show a common trend, reinforcing the hypothesis that no
fast thermal relaxations are responsible of the dynamics of
glycerol, at least in the frequency and temperature regions
covered by Brillouin light scattering. This statement is even
more clear if Fig.~\ref{viscobril} is compared with Fig. 7 of
Ref. \cite{otpfast2}, where $D$($T$) for the Kneser liquid OTP is
reported. In this latter case, a marked temperature dependent
$\alpha/\alpha_{cl}$ ratio has been observed, attributed to the
effects of a thermal relaxation, whose presence has been recently
supported by molecular dynamics simulations \cite{mossa}.

\subsection{Full spectrum analysis}
\label{long}

The traditional analysis, presented in previous section, helps us
to choose an ansatz for the frequency dependent longitudinal
acoustic modulus $M^{\ast}(\omega)$. To this respect, the most
relevant indications we gained from this approach are: {\it i})
at temperatures close to the glass transition the (apparent)
kinematic viscosity obtained from BLS is consistent with the
corresponding unrelaxed values determined by IXS, and {\it ii})
in the high temperature limit the shear viscosity estimated from
BLS collapses on that obtained by viscosimetric measurements.
Both these results indicate the absence of thermal relaxations in
glycerol, in agreement with what inferred some 40 years ago on
the basis of ultrasonic measurements \cite{herzfeld}. Therefore,
in order to model $M^{\ast}(\omega)$, we may suppose that the
structural relaxation is enough to catch the leading
contributions to the relaxational dynamics of glycerol in the GHz
frequency window. Relaxation processes so fast to be out of the
experimental frequency window are described by an instantaneous
term in the modulus function and their effects are taken into
account by the unrelaxed viscosity $\eta_{\infty}$. Following
these indications, our ansatz for the $M^{\ast}(\omega)$ function
consists of a Cole-Davidson function for the $\alpha$ relaxation
plus an instantaneous contribution:

\begin{equation}
M^{\ast}(\omega)=M_{\infty}+\frac{(M_0-M_{\infty})}{(1+i\omega\tau_{\alpha})^{\beta}}
+i\omega \eta_{\infty}.
\end{equation}

Substituting this expression for $M^{\ast}(\omega)$ into Eq.~6,
we have obtained a model function to fit to the BLS spectra. In
particular, this full spectrum analysis has been carried out for
temperatures higher than $T_g$, adopting the following procedure:
(i) The thermal diffusion contribution is considered as a Dirac
$\delta$ term centered at zero frequency, whose amplitude is
proportional to that of the adiabatic part of the spectrum via
the LP-ratio. In fact the characteristic time of the thermal
diffusion mode as measured, for instance, in Ref.~\cite{DiLeo02}
and properly rescaled to the $q$ values accessed to by BLS is
located at about 10 ns, i.e. more than three decades far from the
Brillouin peak in the whole temperature region considered here.
Moreover, even at the highest investigated temperatures, if the
thermal contribution is explicitly considered as a Lorentzian peak
centered at zero frequency, the quality of the fit is unchanged
together with the values of the fitting parameters. (ii) The
values of the relaxed adiabatic sound velocity $c_0$ have been
fixed to those reported in Tab.~\ref{parafit}. (iii)
$\eta_{\infty}/\rho$ has been left to vary in the fit having as
starting point the value determined by IXS
\cite{Sette98,Masciove98,Ruocco99}. The weak T-dependence and the
experimental uncertainty of IXS data discouraged us from fixing
$\eta_{\infty}/\rho$ to a constant value. (iv) The unrelaxed
adiabatic sound velocity $c_{\infty}$ has also been taken as an
adjustable parameter in the whole T-range, while $\tau_{\alpha}$,
and $\beta$ have been considered free or fixed depending on the
temperature region investigated. In particular, in the low
temperature region (T $\leq 255 $ K), being the structural
relaxation out of the Brillouin window, $\tau_{\alpha}$ was fixed
and $\beta$ left free since it affects the shape of the Mountain
region of the spectrum. In this region $\tau_{\alpha}$ has been
imposed to be proportional to the shear viscosity $\eta_S$, i.e.
$\tau_{\alpha}=J \eta_S$, where $J$ is a constant coefficient
having the dimensions of a compliance (see for example
Ref.~\cite{otpfast2}). In contrast, in the highest temperature
region (T $\geq 270 $ K) the structural relaxation becomes
progressively active in the GHz frequency window, so that we are
able to obtain $\tau_{\alpha}$ from the fit, but we need to fix
the value of $\beta$. Its value has been chosen within the limits
suggested from Ref.~\cite{NagelPRA86}, i.e. $0.40\pm 0.05$, and
also consistent with our results at lower T. For some selected
temperatures, namely 295 K, 350 K, and 400 K, we simultaneously
fit BLS ($q$=0.036 nm$^{-1}$) and IXS ($q$=2 nm$^{-1}$) spectra,
following the combined fitting procedure described in
Ref.~\cite{fiorettoPRB02}. In this procedure we use a single IXS
spectrum collected at $q$=2 nm$^{-1}$, since the spectra at
higher $q$'s progressively depart from the hydrodynamic regime.
In the joint-analysis, the values of $c_{\infty}$ and
$\eta_{\infty}$ are mainly related to the position and linewidth
of the Brillouin peaks of the IXS spectra, while $\tau_{\alpha}$
and $\beta$ are mainly related to the shape of BLS spectra. The
obtained best fitting curves are shown in Fig.~\ref{spettri} and
Fig.~\ref{spetIXS} as full lines. The whole set of fitting
parameters is reported in Tab.~\ref{parafit}. The relaxation
times obtained from the BLS data-analysis are also shown in
Fig.~\ref{tauBLS} together with several literature data. In
particular, in the same figure we report three different
evaluations of $\tau_\alpha$ for $T$=350 K: the value obtained
from the described fit procedure (full circle); the value of
$\tau_{max}$ (down triangle) corresponding to the maximum of the
Cole-Davidson function used to model the BLS spectra; and the
characteristic time of longitudinal acoustic mode
$\tau_{LA}=1/2\pi\nu_{LA}$ (star), i.e. the time corresponding to
the maximum of the BLS absorption data in Fig.~\ref{viscocinbril}.
We underline that this last evaluation of $\tau_\alpha$ is model
independent. Fig.~\ref{tauBLS} shows that, at $T$=350 K,
$\tau_{max}$ and $\tau_{LA}$ coincide. This result represents a
strong support of the reliability of the values of the structural
relaxation time as deduced from the full spectrum analysis
presented here.

\begin{table*}
\caption{\label{parafit} Parameters of the full spectrum analysis
of liquid and supercooled glycerol at the temperatures reported in
column 1. Columns 2, 3, and 4 report respectively the
characteristic time of the $\alpha$-CD function, the stretching
parameter, and the average value of $\tau$ determined as
$\langle\tau_{\alpha}\rangle=\beta \tau_{\alpha}$. Columns 5 and
6 report the high and low limiting values of the sound velocity,
respectively. Column 7 reports the instantaneous kinematic
viscosity contribution.}
\begin{ruledtabular}
\begin{tabular}{ccccccc}
$T$  &  $\tau_{\alpha}$    &  $\beta$  &  $<\tau_{\alpha}>$ & $c_{\infty}$ &  $c_{0}$ & $\eta_{\infty}/\rho$\\
$[K]$ &  [10$^{-12}$ s]  &  &  [10$^{-13}$ s]& [m/s] & [m/s] & [10$^{-3}$ cm/s$^{2}$]\\
\hline
& & & & & &  \\
210  &  1.3x10$^9$\footnotemark[1]      &  0.24     &3.1x10$^9$     &  3495& 2088\footnotemark[3] & 2.7\\
225  &  1.9x10$^7$\footnotemark[1]      &  0.27     &5.1x10$^7$     &  3437& 2056\footnotemark[3] & 3.3\\
240  &  6.3x10$^5$\footnotemark[1]      &  0.28     &1.8x10$^6$     &  3372& 2027\footnotemark[3] & 2.7\\
255  &  4.0x10$^4$\footnotemark[1]      &  0.34     &1.4x10$^5$     &  3283& 1996\footnotemark[3] & 5.8\\
275  &  6.0x10$^3$\footnotemark[2]      &  0.35     &2.1x10$^4$     &  3222& 1965\footnotemark[3] & 5.8\\
285  &  1.3x10$^3$\footnotemark[2]      &  0.37     &4.8x10$^3$     &  3160& 1933\footnotemark[3] & 6.1\\
295  &  5.1x10$^{2}$\footnotemark[2]    &  0.37     &1.9x10$^3$     &  3105& 1919\footnotemark[4] & 9.0\\
326  &  7.7x10$^{1}$\footnotemark[2]    &  0.36     &2.8x10$^2$     &  2941& 1848\footnotemark[4] & 11.0\\
333  &  5.4x10$^{1}$\footnotemark[2]    &  0.36     &2.0x10$^2$     &  2894& 1832\footnotemark[4] & 9.0\\
350  &  3.1x10$^{1}$\footnotemark[2]    &  0.36     &1.1x10$^2$     &  2762& 1793\footnotemark[4] & 11.0\\
400  &  1.2x10$^{1}$\footnotemark[2]    &  0.36     &4.1x10$^1$     &  2404& 1675\footnotemark[5] & 11.0\\
\end{tabular}
\end{ruledtabular}
\footnotetext[1]{Fixed on the basis of the temperature behavior of
the shear viscosity.}\footnotetext[2]{Free parameter in the fit.
}\footnotetext[3]{Ref.~\cite{Meis60}.} \footnotetext[4]{Present
work.} \footnotetext[5]{Value obtained from the interpolation of
our data of $c_0$ and those reported in
Ref.~\cite{TandyGrubbs94}.}
\end{table*}

\section{Discussion}
\label{disc}

\subsection{The structural relaxation time and the stretching parameter}
\label{tau}

\begin{figure} { \hspace{-0.0cm}\vspace{-0.8cm}\scalebox{0.80}
{\includegraphics{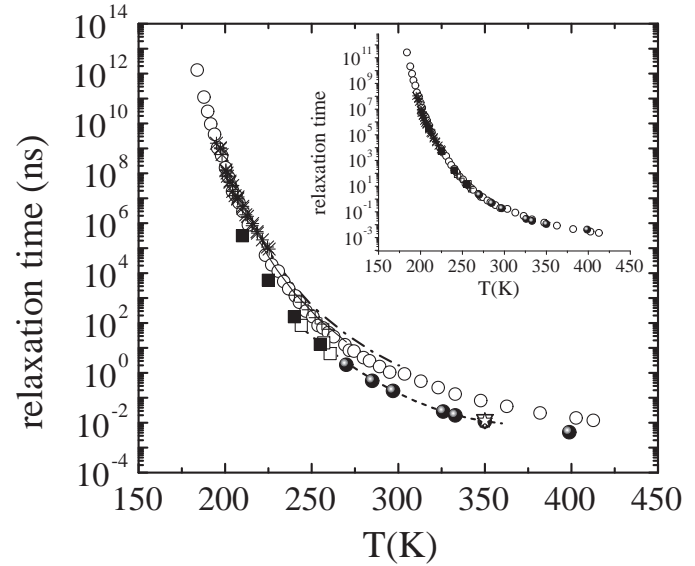}}}\caption{\label{tauBLS} Structural
relaxation time of glycerol from the BLS full spectrum analysis
compared with various literature data. Full circles and squares
represent our BLS values obtained in the fitting procedure by
leaving the $\tau_\alpha$ parameter free or by imposing it to be
proportional to the shear viscosity, respectively. The star
represents the value $\tau_{LA}$ corresponding to the maximum of
the acoustic attenuation for the longitudinal modes; the down
triangle marks the value of $\tau_{\alpha}$ at 350 K
corresponding to the maximum of the Cole-Davidson function used
to model the BLS spectra. The open squares represent ultrasonic
data taken at 2, 10, and 27 MHz \cite{NagelPRA86}, pluses are
from Ref.~\cite{DiLeo02}, crosses are from
Ref.~\cite{Lallemand80}, open circles are from
Ref.~\cite{Schneider98}. The short-dashed line is the polynomial
law given in Ref.~\cite{YanChen88}. The dash-dot line is the
Vogel-Fulcher equation reported in Ref.~\cite{Dux79} describing
photon correlation data. The solid line is the Vogel-Fulcher
equation obtained in Ref.~\cite{Nelson00} from a fit to ISTS
data. In the inset, the same data are reported after rescaling by
a constant. The collapse of all the data on a single master-curve
indicates the validity of the $\alpha$-scale universality in
glycerol over a wide time-temperature region.}
\end{figure}

The temperature behavior of the structural relaxation time
determined in the previous paragraph is compared  with that
obtained by different spectroscopic techniques (dielectric
spectroscopy, impulsive stimulated Brillouin scattering,
impulsive stimulated thermal scattering, ultrasonic techniques,
depolarized light scattering) in Fig.~\ref{tauBLS}. The inset of
the figure shows that the data obtained by these techniques, once
rescaled by a constant, collapse into a single master curve,
indicating that the $\alpha$-scale universality is obeyed in
glycerol in a wide time-temperature range. A simple explanation
for the difference in the absolute values of $\tau$ is the
different microscopic correlators measured by the different
techniques. Next in importance is the definition of the
characteristic time itself. In some cases, like in BLS and
ultrasonic measurements, the longitudinal acoustic modulus
formalism has been used, with a specific CD complex function to
describe the structural relaxation process. In some other case,
like in dielectric spectroscopy measurements, the compliance
formalism was chosen and, although a CD relaxation function was
used, its parameters are not analytically related to those of the
CD modulus function. Moreover, some authors reports the values of
$\tau_\alpha$ directly obtained from the fit
\cite{Dux79,YanChen88,Lallemand80}; others report the average
value $<\tau>$ ($\beta \tau$, in case of a CD function;
($\tau/\beta) \Gamma(1/\beta)$, where $\Gamma$ is the Gamma
function, in the case of a Kolhraush-Williams-Watt function)
\cite{Nelson00,DiLeo02};  others report the value corresponding
to the maximum of the absorption $\tau_{max}$
\cite{NagelPRA86,Schneider98}.

Despite these technical obstacles, one reasonably expects that,
at least in those cases where the same correlator is measured, a
quantitative agreement between relaxation times obtained by
different techniques should be reached. In particular BLS,
ultrasound, ISBS, and ISTS all probe density
fluctuations. The agreement between BLS, ultrasonic, and ISBS
$\tau$-modulus data (present work, \cite{NagelPRA86},
\cite{YanChen88}) is evident in Fig.~\ref{tauBLS}. Different from
this, ISTS $\tau$-compliance data \cite{Nelson00,DiLeo02} are
shifted by about a factor ten towards longer times. Even after a
conversion of the BLS data to the compliance formalism, a shift of
about a factor three persists between the BLS and the ISTS data.
At present, we cannot find any explanation for this interesting
anomaly, and we limit ourselves to suggest to perform this
comparison on other glass-forming systems.

The present full spectrum analysis provides information also on
the T-dependence of the shape parameter $\beta_{CD}$ of the
$\alpha$-relaxation function. In particular, an almost constant
value, namely 0.37$\pm$0.03, was found for temperatures between
270 and 400 K, close to the value 0.40$\pm$0.05 obtained in a
previous ultrasonic investigation \cite{NagelPRA86}. The value of
the stretching parameter decreases with temperature, as shown in
Tab.~\ref{parafit} - the same trend being also found in the
dielectric investigation reported in Ref.~\cite{Schneider98}, the
only one which covers an extremely wide temperature-frequency
range. For a quantitative comparison with the values obtained by other
techniques where compliance relaxation functions are used, like
ISTS, photon correlation spectroscopy, and specific heat
spectroscopy, we converted our modulus-based data into compliance-based
data obtaining the values reported in Tab.~\ref{beta}. Moreover, for
those techniques where the stretching parameter was referred to a
Kohlrausch-Williams-Watt function (KWW), we have transformed
$\beta_{KWW}$ into $\beta_{CD}$ using the equations reported in
Ref.~\cite{lindsey}. Our BLS data compare fairly well with an
average value of $\beta_{CD}$=0.45$\pm$0.05 estimated from ISTS
measurements at different T and exchanged momentum,
$\beta_{CD}$=0.40$\pm$0.05 from PCS data \cite{Demoulin74,Dux79}
and $\beta_{CD}$=0.51$\pm$0.03 from specific heat
spectroscopy data \cite{Birge}.

\begin{table}
\caption{\label{beta} Summary of the shapes of different
compliance functions. }
\begin{ruledtabular}
\begin{tabular}{cccc}
Technique  &  $\beta_{CD}$     &  Temperature range& Refs. \\
\hline
& & &  \\
BLS             &  0.40$\pm$0.05          &210 K       &present work \\
BLS             &  0.40$\pm$0.05          &225 K       &present work \\
BLS             &  0.53$\pm$0.05          &240 K       &present work \\
BLS             &  0.56$\pm$0.05          &255 K       &present work \\
ISTS            &  0.45$\pm$0.05          &200-340 K    &\cite{Nelson00,DiLeo02}\\
PCS             &  0.40$\pm$0.03          &195-225 K     &\cite{Demoulin74,Dux79} \\
Ultrasonics     &  0.42$\pm$0.05          &235-300 K     & \cite{NagelPRA86} \\
Specific heat   &  0.51$\pm$0.03          &195-225 K      &\cite{Birge} \\
Dielectrics     &  0.55$\div$0.65          &195-300 K      &\cite{Schneider98}\\
\end{tabular}
\end{ruledtabular}
\end{table}

\subsection{The limiting high frequency sound velocity} \label{cinfinito}

\begin{figure}\vspace{0.0cm} \hspace{2.2cm}{ \scalebox{0.8}
{\includegraphics{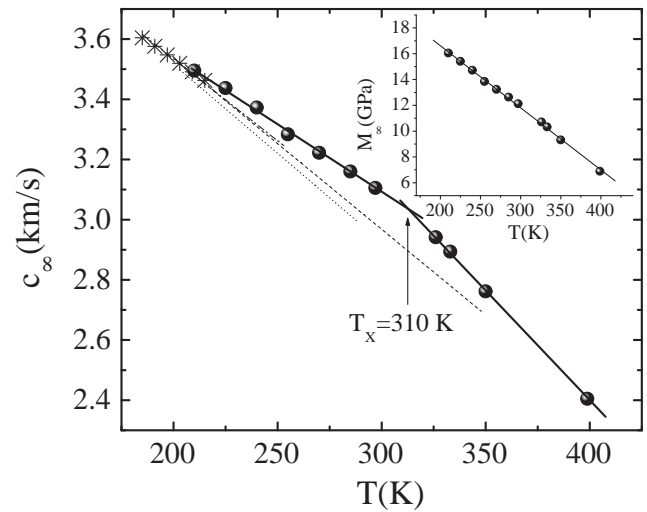}}} \caption{\label{cinf} Unrelaxed
sound velocity. The values of $c_{\infty}$ determined from the
BLS full spectrum analysis (full circles) are compared with
various proposed extrapolations: dashed, dotted, and dash-dot
lines are from Refs. \cite{YanChen88}, \cite{Meis60}, and
\cite{TandyGrubbs94}, respectively. Stars represent data of
$c_{\infty}$ from the linear expression given in
Ref.~\cite{TandyGrubbs94} in the 185.5-218.4 K temperature range.
Solid lines are a guide for the eye. In the inset circles
represent the unrelaxed longitudinal modulus $M_{\infty}=\rho
c_{\infty}^2$. The solid line represents the equation obtained by
a linear fit of the modulus data: $
M_{\infty}$(T)=[26.2$\pm$0.2-0.05$\pm$0.01 T] GPa.}
\end{figure}

The limiting high frequency values of the adiabatic sound
velocity obtained from the fit of the Brillouin spectra are shown
in Fig.~\ref{cinf}. In the same figure, we report two different
ansatzs for the unrelaxed speed of sound obtained assuming either
a linear $T$-dependence of $M_{\infty}$ \cite{Meis60} or a linear
$T$-dependence of $c_{\infty}$ \cite{YanChen88,TandyGrubbs94}. In
Ref.~\cite{TandyGrubbs94}, using stimulated Brillouin gain
spectroscopy, a slight but continuous curvature in the
$c_{\infty}$ data was observed and two different expressions were
reported. Our values of $c_{\infty}$ agree with those given in
Ref.~\cite{TandyGrubbs94} for the 185.5-281.4 K temperature range
(stars in Fig.~\ref{cinf}). However, we observe that an
extrapolation of this equation to higher temperatures is not able
to reproduce our BLS results. In the cases of
Refs.~\cite{YanChen88,Meis60}, the linear expressions reported in
Fig.~\ref{cinf} are derived as extrapolations of the data
collected at low frequencies, in a region where, possibly,
complete unrelaxed conditions are not fulfilled. The values of
$c_{\infty}$ obtained from the present BLS analysis display a
slope smaller than in previous determinations and a decreasing
trend on increasing temperature with a possible change of regime
at about 310 K. This estimation is in remarkable agreement with a
previous IXS determination, namely $T_x=300\pm20$ K
\cite{Masciove98}, which was suggested to mark the microscopic
transition between two different dynamical regimes related to the
glass and liquid phases, respectively.

It is interesting to notice that a different representation can
be given for the same data by plotting the longitudinal modulus
$M_{\infty}=\rho c_{\infty}^2$ as a function of $T$. The inset of
Fig.~\ref{cinf} shows that $M_{\infty}(T)$ can be reasonably
approximated by the linear behavior
$M_{\infty}(T)=[26.2(2)-0.05(1) T]$ GPa. This is similar to what
found for the $T$-dependence of the limiting shear modulus of
glycerol \cite{Scarponi03}. This linear behavior of the unrelaxed
moduli is not universal in supercooled liquids. Indeed, in many
glass-formers, a linear extrapolation leads to unphysical low
values of $M_{\infty}$ in the relaxation region. Rather, the
limiting compliance $J_{\infty}$, the reciprocal of $M_{\infty}$,
is found often to vary linearly with temperature
\cite{Torell,harrison}.

\subsection{The non-ergodicity factor} \label{DW}

\begin{figure}\vspace{0.0cm} \hspace{-2.0cm}{ \scalebox{0.8}
{\includegraphics{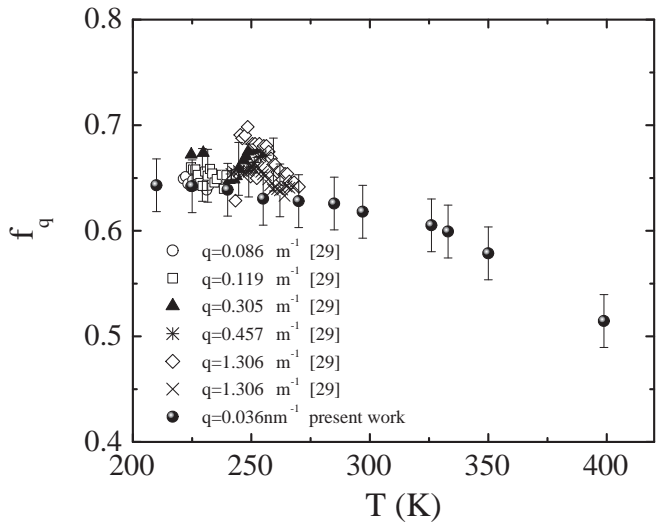}}} \caption{\label{DebyeWaller} The
non-ergodicity factor in glycerol as a function of temperature.
The values of $f_0$ determined from our BLS full spectrum
analysis are compared with the data collected at different
wavevectors and reported in Ref.~\cite{Nelson00}.}
\end{figure}

A detailed analysis in terms of the predictions of the
mode-coupling theory is beyond the subject of this work, we only
present here an estimation of the non ergodicity parameter, $f$,
derived from our isotropic Brillouin spectra. With BLS, in fact,
we directly probe the density-density correlators, and thus we
can extend the temperature interval recently investigated in
Ref.~\cite{Nelson00}. In the zero exchanged momentum limit,
$q\rightarrow 0$, the non-ergodicity factor, $f_q$, can be
calculated  via the equation $f_0=1-c_0^2/c_{\infty}^2$
\cite{goetzeLONG}. This is equivalent to integrate the Mountain
peak in quasi elastic light scattering spectra or the quasi
elastic peak in neutron scattering spectra. Therefore, the $f_0$
parameter has been calculated from the $c_0$ and $c_{\infty}$
values reported in Tab.~\ref{parafit}. The temperature interval
investigated here, namely 210-400 K, includes several previously
different $T_c$ estimations. Looking at Fig.~\ref{DebyeWaller},
it is evident that $f_0$ decreases with increasing temperature,
but no signature of a cusp-like behavior emerges in glycerol.
Fig.~\ref{DebyeWaller} also shows that our results are consistent
with those recently obtained by Paolucci and Nelson
\cite{Nelson00} who measured the non-ergodicity factor from the
strength of the structural relaxation using the ISTS technique in
the 228-268 K temperature range. However, although the existence
of the critical temperature cannot be supported here, we consider
this check only a preliminary test of the MCT predictions, and it
is not possible to exclude the validity of a "non-idealized''
mode-coupling description for glycerol.

\section{Conclusions}
\label{concl}

Brillouin light scattering measurements of glycerol have been
presented, performed in a wide temperature region ranging from
the high temperature liquid to the glassy state. Our investigation
was aimed at understanding the dynamics of density fluctuations
in a prototypical intermediate glass-former in the high frequency
region where the precursor of the glass transition, i.e. the
structural relaxation, is present above $T_g$. The associated
nature of glycerol plays in favor of a simple relaxation pattern,
where the additional effects of internal thermal relaxations is
negligible in the GHz region. Focusing on the narrow frequency
region of the Brillouin peak, the temperature behavior of the
characteristic frequency and lifetime of the longitudinal acoustic
modes has been analyzed. In particular, we have drawn the
attention on the behavior of the apparent kinematic viscosity
that, at temperatures close to and below $T_g$, approaches the
unrelaxed value estimated from IXS data. Different from glass
forming systems that show secondary relaxations of intramolecular
nature, no excess of attenuation for the longitudinal acoustic
modes has been revealed near the glass transition temperature.
Moreover, the longitudinal to shear viscosity ratio is found to
be almost temperature independent in the non-dispersive
high-temperature region. These findings suggest that an
$\alpha$-relaxation only model plus an instantaneous viscosity
term are able to catch the leading contributions to the dynamics
of the density fluctuations. The choice of this model was also
supported by a combined analysis of light and inelastic x-ray
scattering spectra. Starting from this guess, we analyzed
high-resolution BLS spectra over two decades in frequency (full
spectrum analysis) and we were able to obtain details on the
relaxation parameters from the shape of the Brillouin peaks and
of the Mountain region of the spectrum.

The temperature behavior of the structural relaxation time,
$\tau_{\alpha}(T)$, obtained by this fitting procedure, has been
compared with the results coming from different spectroscopic
techniques. The $\alpha$-scale universality is obeyed for the
structural process, i.e. the values of $\tau_\alpha$ are
proportional to those revealed by different experimental
techniques. Moreover, a significant difference of the absolute
values of our $\tau_\alpha$ data is observed with respect to
those obtained by ISTS, a technique which probes the density
fluctuations as well. At the moment, we have no explanation for
this anomaly. It would be interesting to investigate whether this
anomaly holds for other glass-formers as well. Finally,
preliminary considerations have been presented on the predictions
of the mode coupling theory. In particular, in the analysis
presented here, the calculated non-ergodicity factor, $f_0$, is
found to decrease with increasing temperature, but no signature
of a cusp-like behavior is revealed in glycerol. We can thus
argue that the predictions of the idealized MCT are not fulfilled
in this system. Of course, this does not imply that MCT is not
appropriate to describe the dynamics of glycerol. Instead, this
suggests that a more complete version of the theory should be
applied in this case. To this respect, the model introduced by
Franosch et al. \cite{franosch97}, including both linear and
quadratic coupling contributions and tested on depolarized light
scattering spectra of glycerol, seems to be a suitable candidate
to be tested on spectra of density fluctuations. Work along this
line is in progress.

\end{document}